\documentclass[12pt]{article}
\def\lae{\;^{<}_{\sim} \;} \def\gae{\; ^{>}_{\sim} \;}

\title{\textbf{Hilltop Supernatural Inflation and  Gravitino Problem}}

{\author{\\[1cm]
{\sc \large Kazunori Kohri$^{1,2, \dag}$ and Chia-Min Lin$^{3, \ddag}$}\\
{\sl\small $^1$Cosmophysics group, Theory Center, IPNS, KEK, Tsukuba 305-0801, Japan }\\
{\sl\small $^2$Department of Physics, Tohoku University, Sendai 980-8578, Japan }\\
{\sl\small $^3$Department of Physics, National Tsing Hua University, Hsinchu, Taiwan 300 }\\
}}

\usepackage[margin=2cm]{geometry}
\usepackage{graphicx,color}
\usepackage{subfigure}
\begin{document}

\maketitle

\begin{abstract}
In this paper, we explore the parameter space of hilltop supernatural
inflation model and show the regime within which there is no
gravitino problem even if we consider both thermal and nonthermal
production mechanisms. We make plots for the allowed reheating temperature as a function of gravitino mass by constraints from big-bang nucleosynthesis. We also plot the constraint when gravitino is assumed to be stable and plays the role of dark matter.
\end{abstract}


\footnoterule{\small $^\dag$kohri@post.kek.jp,
$^\ddag$cmlin@phys.nthu.edu.tw}

\newpage

\section{Introduction}
Recent WMAP 7-year data \cite{Komatsu:2010fb} suggest a red spectrum of cosmic microwave background (CMB) with $n_s \simeq 0.96$ which supports the idea of hilltop inflation models \cite{Boubekeur:2005zm, Kohri:2007gq} where the inflaton sits near the top of a concave downward potential hill when cosmologically interesting scale exit horizon (i.e. the number of e-folds $N=50 \sim 60$)\footnote{See Ref.~\cite{textbook} for the general review of inflation models.}.

The inflation scale is currently an unknown question. We may be able
to know it if gravitational waves are detected in the near future, for
example, via analysis of B-mode polarization of CMB data from PLANCK
satellite \cite{:2006uk,Komatsu:2009kd}, the ground-based detectors
QUIET+PolarBeaR~\cite{Hazumi:2008zz}, or  KEK's future CMB satellite
experiment, LiteBIRD~\cite{Hazumi:2008zz,LiteBird}. However, for
single-field slow roll inflation we can at least estimate the scale of
inflation via dimensional estimation and it seems the most natural
value of the scale is grand unification theory (GUT) scale. On the
other hand, if we do not restrict ourself by using a single field,
(for example, in the case of hybrid inflation, we use two-fields) the
scale can be lowered. The reason is one field is used to provide the
scalar potential and the other field can have a flatter potential
since the end of inflation is determined by the waterfall field
to become tachyonic. Although the potential energy is mainly from the waterfall field, the curvature perturbation is from the quantum fluctuation of the inflaton field which is slow-rolling during inflation. The potential form for the inflaton
$\Phi$ of a hybrid inflation (during inflation) is given by
\begin{equation}
V(\Phi)=V_0+\frac{1}{2}m^2\Phi^2,
\end{equation}
where $V \simeq V_0$.
The spectrum is
\begin{equation}
P_R=\frac{1}{12\pi^2M_P^6}\frac{V^3}{V'^2} \;,
\end{equation}
where prime denotes derivative with respect to $\Phi$. The spectrum is
restricted to be $P_R^{1/2} \sim 5 \times 10^{-5}$ from
CMB~\cite{Komatsu:2010fb}. We call this CMB normalization in this
paper. As can be seen from the spectrum, we can lower the scale of
inflation $V$ with a small $V^{\prime}$ while fixing the spectrum. An
interesting possibility is to reduce the scale to a supersymmetry
(SUSY) breaking scale. This is the case of supernatural inflation
\cite{Randall:1995dj, Randall:1996ip} in which a gravity mediated SUSY
breaking scale ($V^{1/4} \simeq 10^{11}\mbox{ GeV}$) is
chosen. Interestingly, the model can work without fine-tuning with the
mass $m$ of order TeV which is the typical soft mass in the framework
of SUSY. The characteristic feature of this model is that the spectral
index $n_s$ is predicted to be blue ($n_s>1$), because the potential
is concave upward. However, the recent WMAP data suggests it to be red
($n_s \sim 0.96$)~\cite{Komatsu:2010fb}. It is well-known that hilltop
inflation can produce a red spectrum~\cite{Boubekeur:2005zm,
Kohri:2007gq}, so it is not surprising that if we can convert
supernatural inflation into a hilltop form, the spectral index can be
reduced to fit WMAP data~\footnote{See the type III  hilltop inflation
model proposed in Ref.~\cite{Kohri:2007gq}.}. What interesting is that
there is a way of achieving this without fine-tuning as well
\cite{Lin:2009yt}. We call it hilltop supernatural inflation in this
paper.

People work on SUSY inflation models know that there is a gravitino
problem coming from thermally produced unwanted gravitinos which can
put an upper bound for the reheating temperature. It is now becoming
well-known also that there may be overproduction of unwanted
gravitinos nonthermally by the inflaton decay \cite{Kawasaki:2006gs,
Dine:2006ii, Kawasaki:2006hm, Endo:2006qk, Kawasaki:2006mb,
Endo:2007ih, Takahashi:2007gw} which may put a lower bound for the
reheating temperature. It was shown that many SUSY hybrid inflation
models, for example, F- and D-term inflation suffers from this latter
type of the gravitino problem.~\footnote{For example, see also the discussion in Ref.~\cite{Nakayama:2010xf} and
references therein.} It is shown in \cite{Randall:1995dj} that
supernatural inflation does not have former type of the gravitino
problem. The hilltop version of it does not change the energy scale
nor the waterfall sector therefore the same conclusion applies in both
cases. Now a natural question to ask is whether (hilltop) supernatural
inflation as a SUSY hybrid inflation has this new gravitino
problem. This paper is mainly addressed on this question.

This paper is organized as follows. In Sec.~\ref{sec2}, we review the idea of hilltop supernatural inflation. In Sec.~\ref{sec3}, we briefly summarize (new) gravitino problem. In Sec.~\ref{sec4}, we investigate the allowed reheating temperature as a function of gravitino mass. We consider both constraints from thermally and nonthermally produced gravitinos. We also consider the constraint from gravitino being a dark matter. Sec.~\ref{sec5} is our conclusion.

\section{Hilltop Supernatural Inflation}
\label{sec2}
We consider a hybrid inflation from a flat direction in the framework of SUSY which will play the role of an inflaton. A flat direction is normally lifted by supersymmetry breaking terms and non-renormalizable terms with the superpotential
\begin{equation}
W=\lambda_p\frac{\Phi^p}{pM^{p-3}_P}
\end{equation}
where $p>3$  and $\lambda_p \sim O(1)$. The scalar potential along the flat direction reads (after minimizing the potential along the angular direction)
\begin{equation}
V(\Phi)=\frac{1}{2}m^2\Phi^2-A\frac{\lambda_p \Phi^p}{p M^{p-3}_P}+\lambda_p^2\frac{\Phi^{2(p-1)}}{M^{2(p-3)}_P},
\end{equation}
where the first and second terms on the right-hand side are the soft mass term and the A-term respectively. The last term is simply the F-term potential of the superpotential. For gravity mediation SUSY breaking, we have $m \sim A \sim O(\mbox{TeV})$. We will  focus on the case $p=4$ (smallest $p$) and neglect the last term\footnote{The reason is for our setup the field value of $\Phi$ is small enough. Therefore compare with the second term, the last term can be neglected. See \cite{Lin:2009yt} for the details.}.  We hope to add to this potential a (dominated) constant term $V_0$ during inflation. This can be achieved, for example, by coupling $\Phi$ to a waterfall field $\phi$ via a superpotential of the form \cite{Randall:1995dj}
\begin{equation}
W=\frac{\Phi^2 \phi^2}{2M^{\prime}}
\label{coupling}
\end{equation}
where $M^{\prime}$ is some large mass scale. The potential of the waterfall field (without the above interaction term) has the form
\begin{equation}
V(\phi)=M_S^4 f(\phi/M_P),
\end{equation}
where $M_S$ is the SUSY breaking scale which we choose as $M_S \simeq 10^{11} \mbox{GeV} \simeq 10^{-7} M_P$ (gravity mediation\footnote{Our model can work as well for the case of anomaly and mirage mediation because the scale of SUSY breaking and gravitino mass are similar to gravity mediation. For gauge mediation case, one of the authors has considered the consequences in \cite{Lin:2009ux}.}). This potential form is common in the framework of SUSY. The explicit form of $V(\phi)$ is not very important. One of the possible choices is \cite{Linde:1993cn}
\begin{equation}
V(\phi)=M^4_S \left(\frac{\phi^2}{M_P^2}-1\right)^2
\end{equation}
Another choice may be \cite{Randall:1995dj}
\begin{equation}
V(\phi)=M^4_S \cos^2(\phi/\sqrt{2}M_P).
\end{equation}
During inflation, when the field value of inflaton is large, a large mass is given to the waterfall field from Eq.~(\ref{coupling}). This makes $\phi=0$ and $V_0=M_S^4$. After inflation, the waterfall field rolls down to its vacuum expectation value (VEV) $\sim M_P$. The value of the mass of $\phi$ is around $m_\phi \sim O(\mbox{TeV})$\footnote{This can be estimated for example by $m_\phi^2 \Delta \phi^2 \sim M_S^4$ with $\Delta \phi \sim M_P$. However, we actually require $m_\phi \gae \mbox{TeV}$ in order for hybrid inflation to end promptly once the mass of waterfall field becomes tachyonic, but too much deviation form TeV would be unnatural. It is also possible to have $m_\phi \lae O(\mbox{TeV})$. In this case, a second stage of inflation could occur. We may consider this in our future work.}.

Hence we obtain a SUSY hybrid inflation which we call \textit{hilltop supernatural inflation}.
The potential during inflation is given by \cite{Lin:2009yt}
\begin{eqnarray}
V(\Phi)&=&V_0+\frac{1}{2}m^2\Phi^2-\frac{\lambda_4 A \Phi^4}{4M_P} \nonumber \\
       &\equiv& V_0 \left(1+\frac{1}{2}\eta_0\frac{\Phi^2}{M_P^2}\right)-\lambda \Phi^4
\label{potential}
\end{eqnarray}
with
\begin{equation}
\eta_0 \equiv \frac{m^2 M_P^2}{V_0}  \;\;\; \mbox{and} \;\;\; \lambda \equiv \frac{\lambda_4 A}{4 M_P}.
\end{equation}

The number of e-folds is given by
\begin{equation}
N=M^{-2}_P\int^{\Phi(N)}_{\Phi_{end}}\frac{V}{V'}d\Phi.
\label{efolds}
\end{equation}
From Eq. (\ref{potential}), we can analytically solve the above integral and obtain
\begin{eqnarray}
\left(\frac{\Phi}{M_P}\right)^2&=&\left(\frac{V_0}{M_P^4}\right)
\frac{\eta_0 e^{2N\eta_0}}{\eta_0 x+4 \lambda (e^{2N\eta_0}-1)}\\
x &
\equiv & \left(\frac{V_0}{M_P^4}\right) \left(\frac{M_P}{\Phi_{end}}\right)^2,
\end{eqnarray}
The spectrum and the spectral index are given respectively by
\begin{eqnarray}
P_R&=&\frac{1}{12\pi^2}e^{-2N\eta_0}\frac{[4\lambda(e^{2N\eta_0}-1)+\eta_0 x]^3}{\eta_0^3(\eta_0 x-4\lambda)^2} \label{eq11}\\
n_s&=&1+2\eta_0 \left[1-\frac{12\lambda e^{2N\eta_0}}{\eta_0 x+4\lambda(e^{2N\eta_0}-1)}\right].
\end{eqnarray}

Since we consider gravity mediation, as mentioned, the natural values of soft SUSY breaking terms,
$m$ and $A$,  are $m \sim A \sim
O(\mbox{TeV}) \sim 10^{-15} M_P$. The coupling $\lambda_4$ is of
$O(1)$, which makes
$\lambda \sim O(10^{-15})$. It is interesting that within those natural values, Eq.~(\ref{eq11}) with CMB normalization can be satisfied and we obtain a successful inflation model with $n_s=0.96$ which fits WMAP data very well.

\section{Gravitino Problem}
\label{sec3}

\subsection{Thermal Production}
After inflation, both the inflaton and the waterfall field start to oscillate. Before the reheating, actually the oscillating field can be the mixture of the inflaton field and the waterfall field. However, the lifetime of the inflaton field should be shorter than that of the waterfall field because the waterfall field has a large vev of the order of $M_{P}$, unlike a negligible value of the inflaton field's vev. This would give a large mass to the inflaton via Eq.~(\ref{coupling}). Then the energy density of the oscillation should be dominated by the waterfall field. Reheating happens via the decay of waterfall field through gauge or Yukawa couplings \cite{Randall:1994fr, Randall:1995dj}, therefore the reheating temperature could be higher than $m_\phi$. Since we are considering a SUSY hybrid inflation, there could be constraints from gravitino production. In the following, we will first explain the gravitino problem from thermal production. We will consider nonthermal production in the next section and then the constraint to reheating temperature.

To simply illustrate the problem for the thermally-produced gravitino
\cite{Kawasaki:1994af, Kawasaki:1994sc, Moroi:1995fs, Bolz:2000fu,Kawasaki:2004qu,Pradler:2006qh,Pradler:2006hh,Rychkov:2007uq,Kawasaki:2008qe, Khlopov:1984pf},
let us start from Boltzmann equation for the gravitino number density
$n_{3/2}$
\begin{equation}
\frac{d n_{3/2}}{dt}+3 H n_{3/2} \sim \langle \sigma v \rangle n^2_{\rm STD},
\end{equation}
where $n_{\rm STD}$ is the number density of standard particle whose
scattering produces gravitino. Then $n_{\rm STD}/s \sim O(1/g_{*})\sim
O(10^{-3})$  with $g_{*} \sim 200$ being an effective number of
relativistic degree of freedom in the particle content of the minimal
SUSY standard model (MSSM). The gravitino number density $\Delta
n_{3/2}$ produced is obtained via solving the Boltzmann equation. We
can approximately estimate the solution as
\begin{equation}
\Delta n_{3/2} \sim \frac{\langle \sigma v \rangle n_{\rm STD}^2}{H},
\end{equation}
therefore
\begin{equation}
Y_{3/2} \equiv \frac{n_{3/2}}{s} \sim \frac{\Delta n_{3/2}}{s}
\sim \frac{\langle \sigma v \rangle n_{\rm STD}}{g_{*}H} \sim \frac1{g_{*}^{3/2}}\frac{T_R}{M_P},
\end{equation}
where we have used $\langle \sigma v \rangle \sim 1/M^2_P$ for massive
gravitino.~\footnote{For simplicity, we are assuming a case of
gravitino mass $m_{3/2} \gae  m_{\tilde{g}}$ with $m_{\tilde{g}}$ to
be  gluino mass.} Therefore thermal production of gravitino abundance
is \emph{proportional} to $T_R$.  A more accurate solution to the
Boltzmann equation for thermal production can be found in
\cite{Bolz:2000fu,Kawasaki:2004qu,Pradler:2006qh,Pradler:2006hh,Rychkov:2007uq,
Kawasaki:2008qe},
which approximately gives
\begin{equation}
Y_{3/2} \simeq 2\times 10^{-16} \times \left(\frac{T_R}{10^{6}\mbox{ GeV}}\right)
\label{eq3}
\end{equation}
Since a large $Y_{3/2}$ conflicts with big-bang nucleosynthesis (BBN),
thermally produced gravitino provide an \emph{upper bound} for the
allowed reheating
temperature~\cite{Kawasaki:2004qu,Kawasaki:2008qe,Kawasaki:2004yh,Kohri:2005wn,Jedamzik:2006xz}.

\subsection{Nonthermal Production}
To illustrate nonthermal production of gravitinos, let us assume our
waterfall field $\phi$ with number density $n_\phi$ decays into two
gravitinos.
\begin{equation}
\phi \rightarrow 2 \psi_{3/2}.
\end{equation}
The number density of gravitino $n_{3/2}$ produced is hence given by
\begin{equation}
n_{3/2}=2 n_\phi B_{3/2},
\end{equation}
where
\begin{equation}
B_{3/2} \equiv \frac{\Gamma_{\phi \rightarrow 2\psi_{3/2}}}{\Gamma_\phi}
\end{equation}
is the branching ratio~\cite{Endo:2006zj,Nakamura:2006uc,Dine:2006ii,Endo:2006tf}. The
waterfall field decays when
\begin{equation}
\Gamma_\phi=H \sim \frac{T^2_R}{M_P}
\label{eq1}
\end{equation}
Therefore
\begin{equation}
Y_{3/2}=2 B_{3/2} \frac{n_\phi}{s} \simeq
\frac{3}{2}\frac{M_P}{m_\phi}\frac{\Gamma_{\phi \rightarrow 2\psi_{3/2}}}{T_R},
\label{eq2}
\end{equation}
where we have used Eq. (\ref{eq1}) and assume the entropy of the universe is from the waterfall field decay. As we can see from Eq. (\ref{eq2}), in the case of nonthermal production of gravitino is \emph{inversely proportional} to the reheating temperature. In our model, $m_\phi<\sqrt{m_{3/2}M_P}$, therefore \cite{Endo:2006zj, Nakamura:2006uc}
\begin{equation}
\Gamma_{\phi \rightarrow 2\psi_{3/2}} \simeq \frac{1}{32\pi} \left(\frac{\langle\phi\rangle}{M_P}\right)^2 \frac{m^3_\phi}{M_P^2}.
\end{equation}
By using Eq. (\ref{eq2}), we obtain \cite{Endo:2007sz}
\begin{equation}
Y_{3/2} \simeq 10^{-17} \left(\frac{T_R}{10^3\mbox{ GeV}}\right)^{-1} \left(\frac{\langle\phi\rangle}{10^{18}\mbox{ GeV}}\right)^2 \left(\frac{m_\phi}{10^{}\mbox{ TeV}}\right)^2
\label{eq4}
\end{equation}
Since a large $Y_{3/2}$ destroys BBN~\cite{Kawasaki:2004yh,Kawasaki:2004qu,Jedamzik:2006xz}, nonthermal production of
gravitino provides a \emph{lower bound} for the reheating temperature.
\section{Reheating Temperature}
\label{sec4}
\begin{figure}[htbp]
\begin{center}
\includegraphics[width=0.5\textwidth]{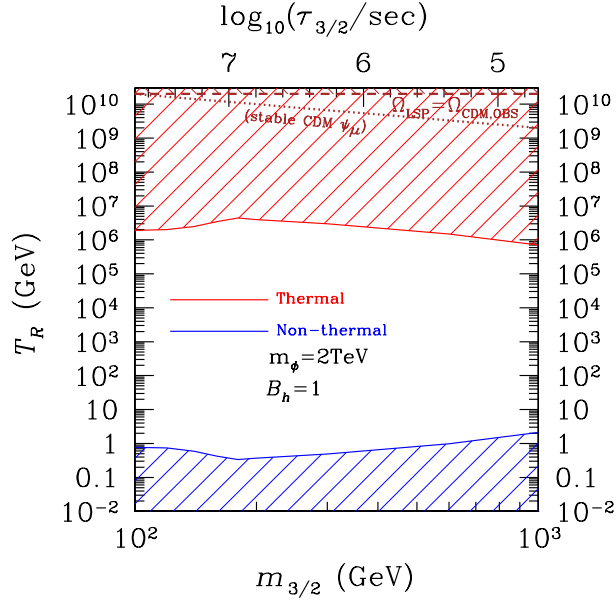}
\caption{Allowed  region in parameter space of $T_R$ versus $m_{3/2}$ with $m_{\phi}=2\mbox{ TeV}$. Note that the constraint can be much
milder only at around $m_{\phi} \sim 2 m_{3/2}$ because of the
suppression of the mode decaying into two gravitinos.}
\label{fig1}
\end{center}
\end{figure}

\begin{figure}[htbp]
\begin{center}
    \includegraphics[width=0.5\textwidth]{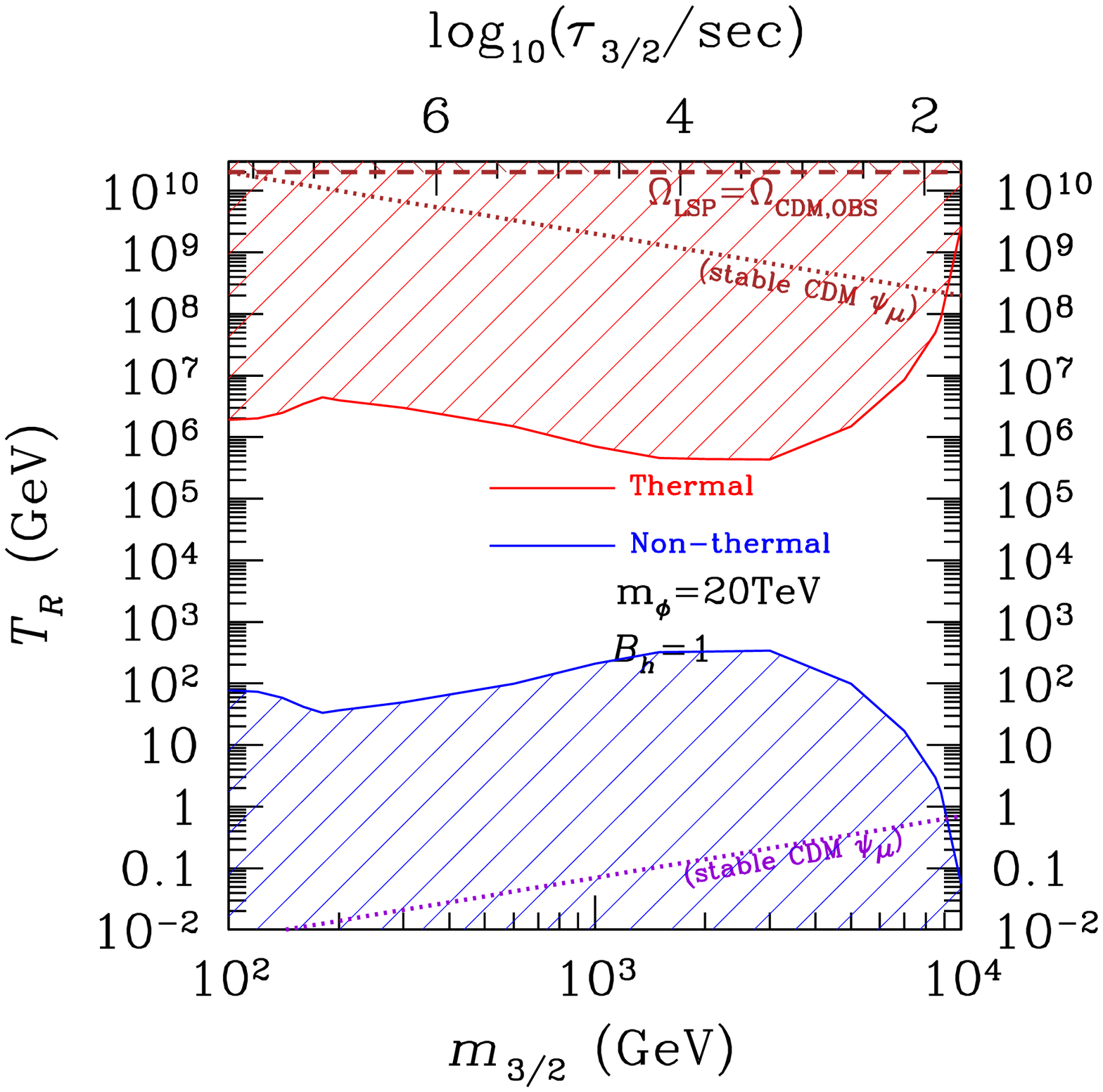}
\caption{$T_R$ versus $m_{3/2}$ with $m_{\phi}=20\mbox{ TeV}$}
\label{fig2}
\end{center}
\end{figure}

\begin{figure}[htbp]
\begin{center}
\includegraphics[width=0.5\textwidth]{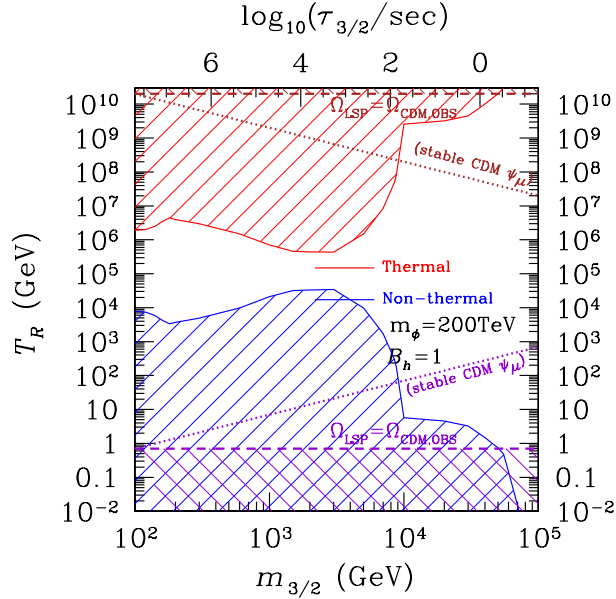}
\caption{$T_R$ versus $m_{3/2}$ with $m_{\phi}=200\mbox{ TeV}$}
\label{fig3}
\end{center}
\end{figure}

Big-bang nucleosynthesis (BBN) put severe constraint on $Y_{3/2}$ (and
hence $T_R$) \cite{Kawasaki:2004yh,Kawasaki:2004qu,
Kohri:2005wn,Jedamzik:2006xz,Kawasaki:2008qe}. The constraint of
$Y_{3/2}$ is roughly $Y_{3/2} \lae 10^{-17}$. From Eqs. (\ref{eq3})
and (\ref{eq4}), by using $\langle\phi\rangle \sim M_P$ and $m_\phi
\gae O(1)\mbox{ TeV}$, we plot the constraint of reheating temperature
as a function of gravitino mass in Figs. \ref{fig1}, \ref{fig2} and
\ref{fig3}. Here we assumed that the hadronic branching ratio is1
$B_h\equiv \Gamma_{\psi_{3/2}\rightarrow
\mbox{hadrons}}/\Gamma_{\psi_{3/2}} \sim 1$ $(\sim 100\%)$ which is natural in
massive unstable gravitino scenario. The dashed line represents the
observational bound on the energy density of the cold dark matter
(CDM) ($\Omega_{\rm CDM}h^2 \lae 0.1$ reported by WMAP
\cite{Komatsu:2010fb}) when a gravitino decays into a Lightest SUSY
Particle (LSP) with the LSP mass $100\mbox{ GeV}$.  Because we did not
consider the thermal relic component of the LSP, this gives a
conservative bound. When we change the
mass of LSP, the constraint can be also changed and scaled accordingly.

We also plot complementary constraints by dotted lines when gravitino is stable and becomes CDM for comparing by using
\begin{equation}
Y_{\rm CDM}=4 \times 10^{-12} \left(\frac{m_{\rm CDM}}{10^{2}\mbox{
  GeV}}\right)^{-1} \left(\frac{\Omega_{\rm CDM} h^2}{0.1}\right).
\end{equation}
This may be unnatural when gravitino mass is much larger than TeV.

\section{Conclusions}
\label{sec5}
In this paper, we have investigate the allowed regime of reheating
temperature as a function of gravitino mass for hilltop supernatural
inflation. We consider both constraints from thermally and
nonthermally produced gravitino and also in the case when gravitino
could become the dark matter. It is not easy to build a SUSY inflation model which
requires no fine-tuning of parameters, predict $n_s=0.96$, and without
gravitino problem. Here we have shown that hilltop supernatural
inflation can meet all these requirement. 

There are some recent works about the
effects of waterfall field to primordial curvature perturbation \cite{Mulryne:2009ci,Lyth:2010ch, Fonseca:2010nk, Abolhasani:2010kr}. Those effects are subdominant and our result is not affected although it may be interesting to investigate them as our future work.

\section*{Acknowledgement}
This work was supported in part by the NSC under grant No. NSC
99-2811-M-007-068, by the NCTS, and by the Boost Program of
NTHU. K.K. was partly supported by the Center for the Promotion of
Integrated Sciences (CPIS) of Sokendai, and Grant-in-Aid for
Scientific Research on Priority Areas No. 18071001,  Scientific
Research (A) No.22244030 and  Innovative Areas No.  21111006.

\end{document}